\documentclass[aip,cha]{revtex4-1}
\usepackage{amsmath,amsbsy,amsfonts,amssymb}
\usepackage[utf8x]{inputenc}
\usepackage{graphicx}
\usepackage{wasysym}
\usepackage{fancyhdr}
\usepackage{hyperref}
\usepackage{color}
\usepackage{float}
\usepackage{enumitem}
\usepackage{subfloat}

\usepackage{dcolumn}
\usepackage{bm}

\begin{document}


\title{Combination anti-coronavirus therapies based on  nonlinear mathematical models}

\author{J. A. Gonz\'alez}
\email{jorgalbert3047@hotmail.com}
\affiliation{ Department of Physics, Florida International University, Miami, Florida 33199, USA}
 
\author{ Z. Akhtar}
\affiliation{Department of Biology, College of Arts and Sciences, University of Miami, Coral Gables, Florida 33146, USA}

\author{D. Andrews}
\affiliation{Medical Campus, Miami Dade College, 950 NW 20th Street, Miami, Florida 33127, USA}

\author{S. Jimenez}
\affiliation{Departamento de Matem\'atica Aplicada a las TT.II, E.T.S.I Telecomunicaci\'on, Universidad Politecnica de Madrid,	28040-Madrid, Spain}

\author{L. Maldonado}
\affiliation{Department of Biological Sciences, Florida International University, Miami, Florida 33199, USA}

\author{T. Oceguera}
\affiliation{Department of Physics, University of Guadalajara, Guadalajara, Jalisco, M\'exico C. P. 44430 } 

\author{I. Rond\'on}
\email{irondon@kias.re.kr}
\affiliation{School of Computational Sciences, Korea Institute for Advanced Study, Seoul 0245, Republic of Korea,\bf{}}

\author{O. Sotolongo-Costa}
\affiliation{Universidad Aut\'onoma del Estado de Morelos,
Cuernavaca, M\'exico, C.P. 62209}

\date{\today}


\begin{abstract}
Using nonlinear mathematical models and experimental data from laboratory and clinical studies, we have designed new combination therapies against COVID-19.
\end{abstract}

\maketitle

\begin{quotation}
Currently there are no approved treatments for the SARS-CoV-2 infection. Moreover, scientists do not know any treatment that would consistently cure COVID-19 patients. This paper is an argument for combination therapies against COVID-19. We investigate a nonlinear dynamical system that describes the SARS-CoV-2 dynamics under the influence of immunological activity and therapy. Using the nonlinear mathematical model and experimental data from laboratory and clinical studies, we have designed new combination therapies against COVID-19. The therapies are based on antivirals in combination with other therapeutic approaches. The general therapeutic plan is the following: Gene therapy and/or Antivirals plus Immunotherapy and Anti-inflammatory drugs and/or drugs that control Cytokine Storms plus Cytotoxic therapies. We believe that these new therapies can improve patient outcomes.
\end{quotation}

\section{Introduction}
\noindent
Emerging viral diseases have caused significant global devastating pandemics, epidemics, and outbreaks (Smallpox, HIV, Polio, 1918 influenza, SARS-CoV, MERS-CoV, Ebola, and SARS-CoV-2).
\\
Currently there are no approved treatments for any human coronavirus infection. Moreover, scientists do not know any treatment that would consistently cure COVID-19 patients. The world is facing a general catastrophe as people see the reality of alarming rises in infections, a building economic crisis, a shortage of ventilators, the lack of coronavirus testing, and many other disasters. The governments are desperate to find a solution. In some cases, they are even promoting unproven “remedies”. The novel coronavirus presents an unprecedented challenge for everybody, including the scientist: the speed at which the virus spreads means they must accelerate their research. We need a treatment that is 95$\%$ effective in order to safely open the countries and save the world from an economic catastrophe.
\\
There is a wealth of literature dedicated to mathematical modeling of the virus-immune-system interaction. See, for example \cite{Ref1}.
There are many treatments in development. However, most of them have drawbacks \cite{Ref2}.
\\
This paper is an argument for combination therapies against COVID-19. We have shown before that combination therapies can be better than monotherapies. For instance, for some cancer tumors, the immunotherapies do not work at all
\cite{Ref3,Ref4,Ref5}.  We have proposed to use a combination of therapies that could eradicate the cancer completely \cite{Ref3,Ref4,Ref5}.
In the present paper we will design new therapies based on antiviral agents in combination with other therapeutic approaches. These new therapies should improve patient outcomes.

\section{Growth models}
\noindent
There are several famous equations that have been used to describe cell population growth: exponential, Gompertz, logistic, and power-law equations \cite{Ref6}.
\\
\\
In reference \cite{Ref7}, a biophysical justification for the Gompertz's equation was presented. 
\\
\\
The deduction is based on the concept of entropy. The entropy definition used in Ref \cite{Ref7} is the well-known Boltzmann-Gibbs extensive entropy. Gonzalez et al. have used the new non-extensive entropy \cite{Ref8,Ref9,Ref10} in the derivation of a new very general growth model \cite{Ref6}. 
\\
\\
The exponential, logistic, Gompertz, and power laws are particular cases of the new equation. The new model has the potential to describe all the known and future experimental data \cite{Ref6}.
\\
The non-extensive parameter $\bm{q}$ \cite{Ref8,Ref9,Ref10,Ref11}  plays an important role in the new model.
Suppose we are studying virus population dynamics.
\\
\\
Different types of viral infections
 can possess different values of the non-extensive para\-meter $\bm{q}$.
\\
Boltzmann-Gibbs statistics satisfactorily describes nature if the microscopic interactions are short-range and the effective microscopic memory is short-ranged, and the boundary conditions are nonfractal.
\\
There is a large series of recently found natural systems that present anomalies that violate the standard Boltzmann-Gibbs method.
\\
A non-extensive thermostatistics, which contains the Boltzmann-Gibbs as a particle case, was proposed in a series of papers \cite{Ref8,Ref9,Ref10,Ref11}.
\\
Nowadays, scientists have produced a large amount of successful applications of the new theory .
These are mostly phenomena in complex systems. \\
The mentioned thermodynamic theory contains the non-extensive entropy:
\begin{align}
S_q = k \frac{1-\sum_{i=1}^{w} p_{i}^{q}}{q-1},
\end{align}
where $\bm{k}$ is a positive constant, $\bm{w}$ is the total number of possibilities of the system,
\begin{align}
	\sum_{i=1}^{w} p_i =1 , \hspace{0.5 cm} q \in \mathbb R,
\end{align}
This expression recovers the Boltzmann-Gibbs entropy, 
$\bm{S_1 = -k \sum_{i=1}^{w} p_i \ln p_i}$, in the limit 
$\bm{q} \rightarrow 1$. Parameter $\bm{q}$ characterizes the degree of non-extensivity of the system. \\
This can be seen in the following rule:
\begin{align}
	S_q(A+B)/k = \left[ S_q (A)/k\right]+ \left[ S_q (B)/k\right] + (1-q)\left[ S_q (A)/k\right]\left[ S_q (B)/k\right],
\end{align}
where $\bm{A}$ and $\bm{B}$ are two independent systems in the sense that 
$\bm{P_{i,j}(A + B) = P_i (A) P_j (B)}$. We could say that parameter $\bm{(1 – q)}$ characterizes the complexity of the system.\\
The case $\bm{1-q  \geq 0}$ implies that the system is resilient. 
\\
For example, this condition indicates that the virus infection will lead to a drug-resistant disease.  In particular, this disease can become resistant to the attack of the immune system and conventional therapy.
\\
\\
The new generalized equation for population growth is the following

\begin{equation}
\label{eq:principal_model}
\frac{dX}{dt} = \frac{k X_\infty}{q-1} \left[ 1 -\left(\frac{X}{X_\infty}\right)^q - \left(1 - \frac{X}{X_\infty}\right)^q\right],
\end{equation}	
where $\bm{X(t)}$ is the growing population, $k$ is certain free parameter, and $\bm{X_\infty}$ is the asymptotic value of $\bm{X(t)}$ when $\bm{t \rightarrow \infty}$.
\\
We have already remarked that this is a very general model that contains most known growth models \cite{Ref6}.\\
\addtocounter{equation}{-1}
\begin{subequations}
	\\	
	Now, we will show that this model is universal in the sense discussed in Ref. \cite{Ref12,Ref13} and includes many others classes of models as particulars cases. \cite{Ref13_1}\\
	For early stages of the infection, \eqref{eq:principal_model} can be written in the following form \cite{Ref13_1}	
	\begin{align}
	\label{eq:model_1a}
	\frac{dX}{dt} = \alpha_q \left( \frac{X}{X_\infty(q)}\right)^q 
	\left[ 1 - \left(\frac{X}{X_\infty(q)} \right)^{1-q}\right],
	\end{align}	
	where $\bm{\alpha_q= \frac{k q X_\infty (q)}{1-q}}$,  $\bm{ X_\infty(q) = q^{\frac{1}{q-1}} X_\infty}$.\\
	An analytical solution to equation \eqref{eq:model_1a} can be expressed as
	\begin{align}
	\label{eq:model_1b}
	\left( \frac{X}{X_\infty (q)}\right)^{1-q} = 1 - \left[ 1 - \left( \frac{X_0}{X_\infty (q)} \right)^{1-q}\right] e^{-q k t}
	\end{align}
	where $\bm{X_0}$ is the initial condition so that $\bm{X(t=0)=X_0}$.\\
		We can re-write solution \eqref{eq:model_1b} as	
		\begin{align}
			\label{eq:model_1c}
			r(\tau)= 1 - e^{-q \tau}
		\end{align} 
		where 
		$\bm{r=\left(\frac{X}{X_\infty(q)} \right)^{1-q}}$ and  $\bm{\tau= kt -\ln(1-r_0)^{1/q}}$.	
\\	
This calculation shows that our model represents a universal growth law \cite{Ref12,Ref13,Ref13_1}.
	\\
	So even this general class of growth laws are a particular case of equation \eqref{eq:principal_model}.
\end{subequations}
\begin{figure} 
	\centering
	\includegraphics[scale=0.8]{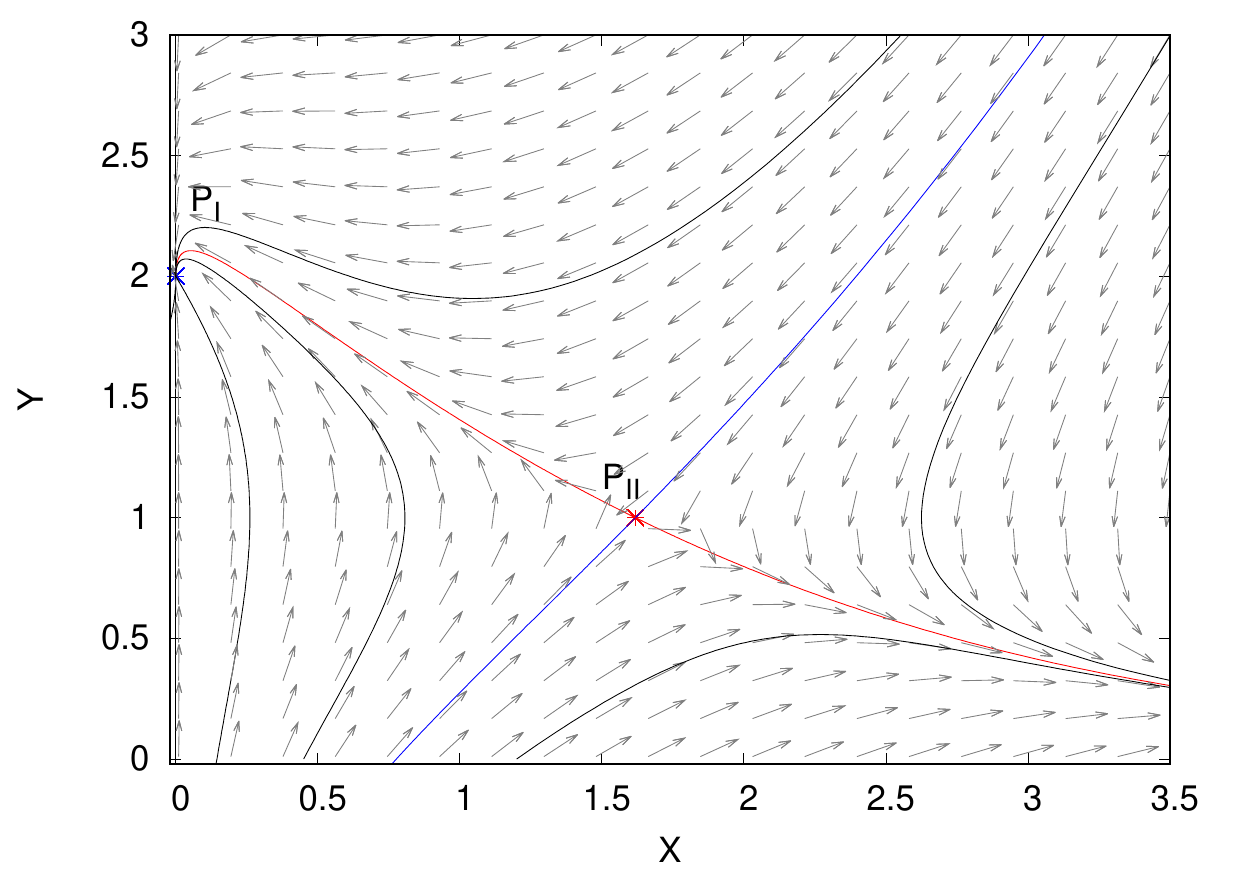}
	\caption{Phase space representation of the dynamics of the dynamical system \eqref{equation2} , \eqref{equation3}. In this case, the fixed point $\mathbf{P_{I}}$ is a stable node, the point $\mathbf{P_{II}}$ is a saddle, and the point where $\mathbf{X \approx X_\infty} $ is a stable node (not shown). The blue line is the stable manifold of the saddle point $\mathbf{P_{II}}$. And the red line is the unstable manifold of the mentioned saddle point. The blue line is a global separatrix of the dynamics. All initial conditions that are on the “left” of the blue line will lead to a phase trajectory that tends to a point where $\mathbf{X=0}$. All initial conditions that are on the “right” of the separatrix will lead to a phase trajectory that tends to the point where $\mathbf{X\approx  X_\infty}$. This picture occurs when $\mathbf{q >1}$ , $\mathbf{bV > af}$.} 
	\label{phasevirus_fig1}
\end{figure}

\section{The model}
\label{section3}
In the present paper, we will investigate the following dynamical system

\begin{align}
\label{equation2}
\frac{dX}{dt}&= \frac{k X_\infty}{q-1}\left[1 - \left(\frac{X}{X_\infty}\right)^q - \left( 1 - \frac{X}{X_\infty}\right)^{q}\right] -bXY  - c_1(t) X,\\
\frac{dY}{dt}& = d(X - eX^2) Y - fY + V  - c_2(t)Y,
\label{equation3}
\end{align}
where $\bm{X}$ denotes the virus population and $\bm{Y}$ denotes the population of lymphocytes. Equation \eqref{equation2} describes the reproduction of the virus. The virus is killed when it meets agents of the immune system (term $\bm{- bXY}$).
\\
\\
The reproduction of the agents of the immune system is described by the term $\bm{d(X-eX^2)}$, where initially the presence of the virus stimulates the reproduction of $\bm{Y(t)}$. When virus load is very large, the person is so sick that the reproduction of $\bm{Y(t)}$ is inhibited. The term $\bm{-fY}$  corresponds to the natural death of lymphocytes. The term $\bm{V}$ represents an external flow of lymphocytes. 
\\
\\
The term $\bm{-c_1(t)X}$ stands for virus-killing process due to different therapies. The term $\bm{-c_2(t)Y}$ shows that therapies can also affect other normal cells (including the immune system).
\\
\\
The system \eqref{equation2} and \eqref{equation3} is inspired by models of the immune system developed in references \cite{Ref14} and \cite{Ref15}.
However, instead of the exponential growth assumed in \cite{Ref14,Ref15}, we are using our growth model given by Eq. \eqref{eq:principal_model}.
\begin{figure} 
	\centering
	\includegraphics[scale=0.8]{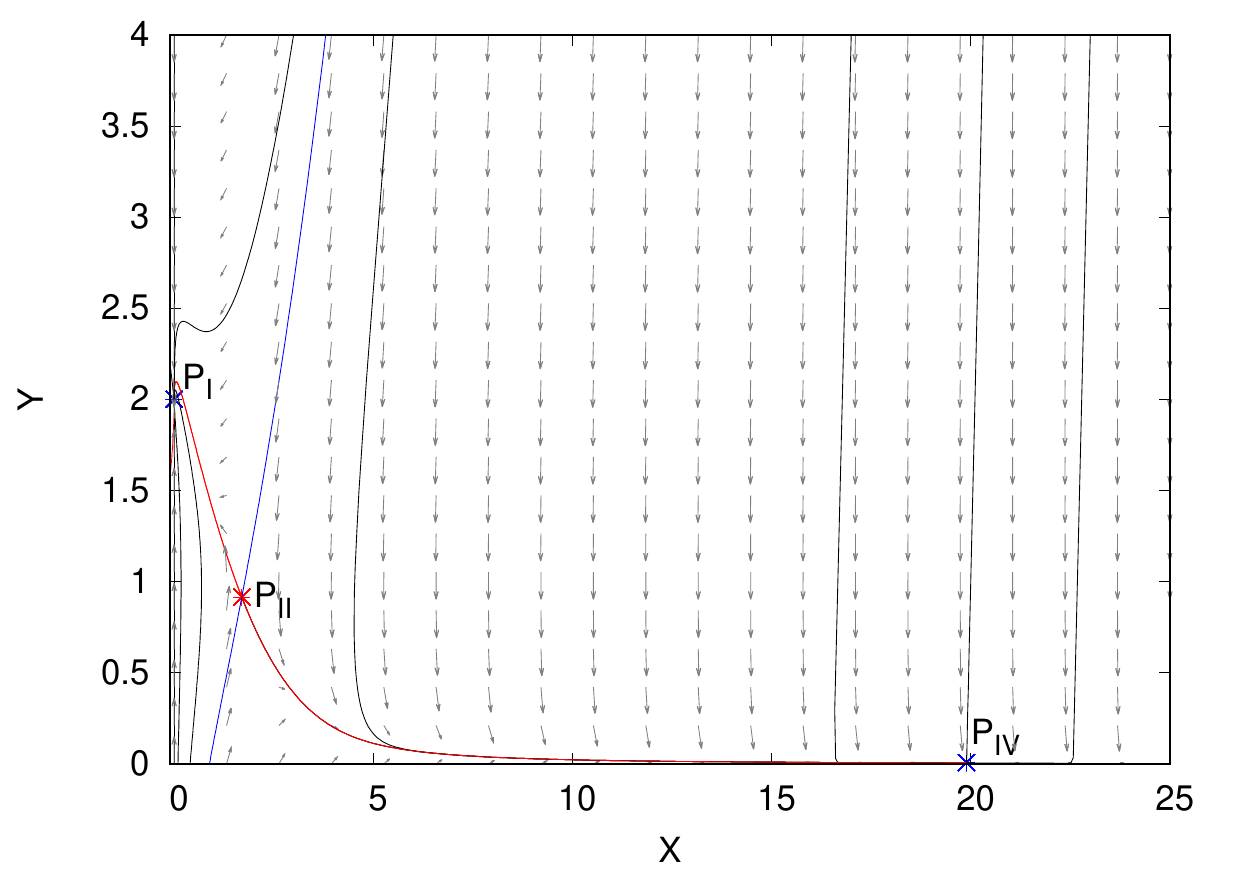}
	\caption{ This situation is topologically equivalent to that shown in  Fig. \eqref{phasevirus_fig1} 
		However, now we can see the fixed point where $\mathbf{X \approx X_\infty}$. (Far right). } 
	\label{phasevirus_fig2}
\end{figure}

\section{Investigation of the model}
\noindent
First, we will consider the case where $\bm{q>1}$, $\bm{X_\infty e>>1}$,
$\bm{c_1(t)=0}$,$\bm{c_2(t)=0}$ .\\
Let us define $\bm{a=\frac{qk}{q-1}}$. 
The dynamical system \eqref{equation2}-\eqref{equation3} can have, in principle, four fixed points
\begin{equation}
\label{equation4}
P_{I}=(X_1,Y_1)= (0,\frac{V}{f})
\end{equation}

\begin{align}
\label{equation5}
P_{II}=(X_2,Y_2),
\end{align}
where $\bm{\frac{1}{2e}<X_2<X_\infty}$, $Y_2=\frac{a}{b}$,
\begin{align}
\label{equation6}
P_{III}=(X_3,Y_3),
\end{align}
where $\bm{0<X_3< \frac{1}{2e}}$, $\bm{Y_3=\frac{a}{b}}$,
\begin{align}
\label{equation7}
P_{IV}=(X_4,Y_4),
\end{align}
where $\bm{X_4 = X_\infty}$,
\\
The conditions for the existence of points $\bm{P_{II}}$ and $\bm{P_{III}}$   are the following inequalities
\begin{align}
\label{equation8}
\frac{1}{4e^2} -h &>  0,
\\
h  >0,
\label{equation9}
\end{align}
where $\bm{h= \left( f\frac{a}{b}- V\right)\frac{b}{e a d} }$
\\
The eigenvalues of the Jacobian matrix corresponding to the fixed point $\bm{P_{I}}$  are
\begin{align}
\label{equation10}
\lambda_{1}^{(I)} &=a -\frac{bV}{f},\\
\label{equation11}
\lambda_{2}^{(I)}&= -f.
\end{align}
If $\bm{af<Vb}$, the fixed point $\bm{P_{I}}$ is a stable node and the fixed point $\bm{P_{II}}$ is a saddle  (See figures  \eqref{phasevirus_fig1} -\eqref{phasevirus_fig2}) . If $\bm{af>Vb}$, and $\bm{h -\frac{1}{4e^2} <0}$, then the four fixed points exist and are non-negative. Both fixed points $\bm{P_{I}}$ and $\bm{P_{II}}$   are now saddles. Between these two points, there is the point $\bm{P_{III}}$, which is stable (See Fig. \eqref{phasevirus_fig3} and Fig.\eqref{phasevirus_fig4})).
\\
\\
If $\bm{af>Vb}$, and $\bm{h - \frac{1}{4e^2}>0}$, then there are only two fixed points: point $\bm{P_{I}}$  which is now unstable and point $\bm{P_{IV}}$, which is stable. As a result, most trajectories tend to point $\bm{P_{IV}}$ (with maximum virus population)  . This is not a very favorable situation for the patient. (See Fig. \eqref{phasevirus_fig5})
\\
\\
In the neighborhood of point  $\bm{P_{II}}$, the separatrix of the saddle can be approximated by the straight line
\begin{align}
\label{equation12}
Y= - \left(\frac{\lambda_{2}^{(II)}}{bX_2}\right) X + \frac{a+\lambda_{2}^{II}}{b}
\end{align}
Any point corresponding to initial conditions of the Cauchy problem on the right of the separatrix leads to a dynamics where the trajectory approaches the point of maximum virus load (point $\bm{P_{IV}}$).\\
On the other hand, if the initial  conditions correspond to a point located on the left of the separatrix, the system will evolve to a stable fixed point.
\\
Using \eqref{equation12}, we can calculate the threshold or critical virus population that would lead to a dynamics approaching point $\bm{P_{IV}}$:

\begin{figure} 
	\centering
	\includegraphics[scale=0.8]{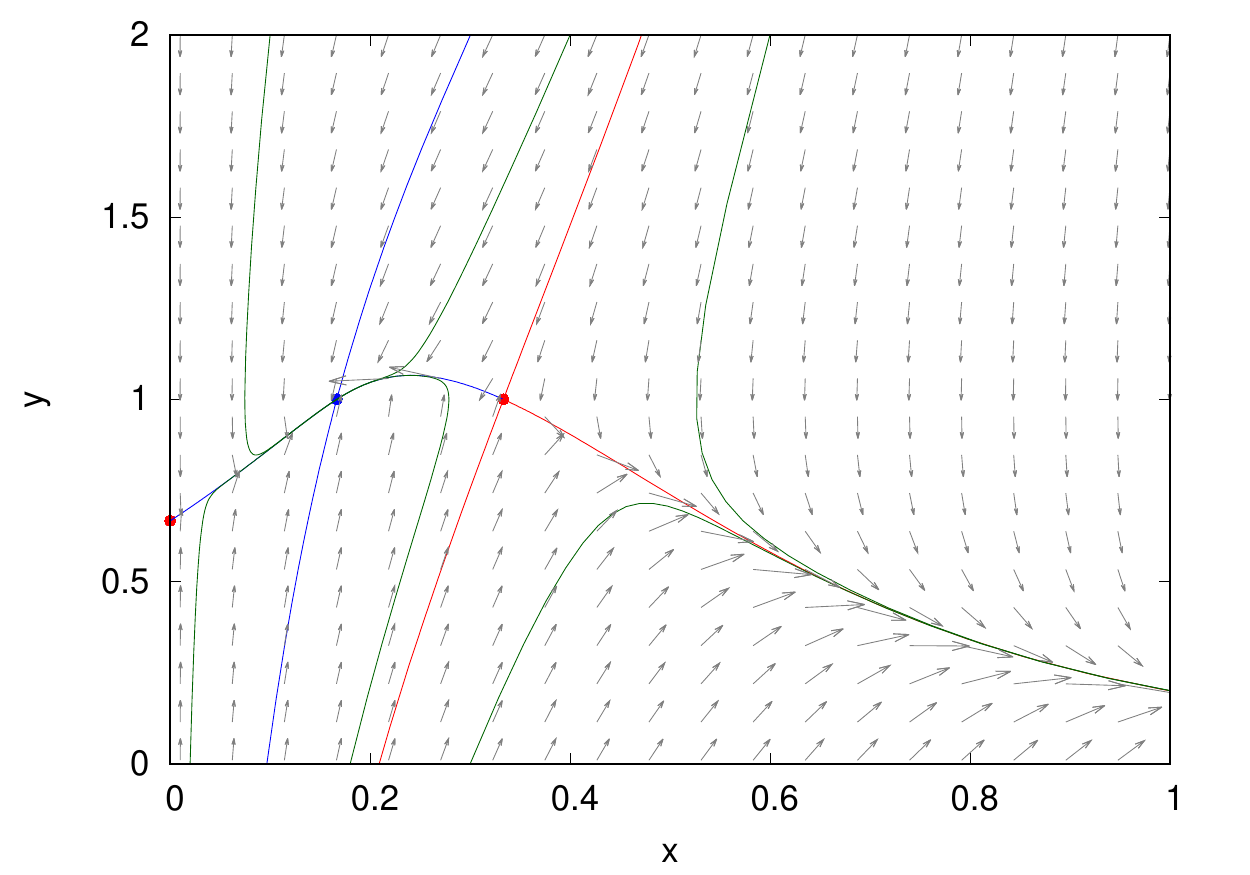}
	\caption{Phase space representation of the dynamics of system \eqref{equation2} ,\eqref{equation3} . In this case, the fixed point $\mathbf{P_{I}} $  is unstable, the fixed point $\mathbf{P_{II}}$  is still a saddle. Now there is a new fixed point $\mathbf{P_{III}} $  that is a stable node (for which $\mathbf{X>0} $). Now the separatrix is represented with a red line. All the phase trajectories that are on the “left” of the separatrix are approaching the point $\mathbf{P_{III}} $ (where $\mathbf{X>0} $). All the phase trajectories that are on the “right” of the separatrix are approaching the fixed point where 
		$\mathbf{X \approx X_\infty}$. Note that there are trajectories for which $\mathbf{X(t)} $  is monotonically increasing from a small value until it reaches its maximum (Point $\mathbf{P_{III}}$ ). There are other trajectories for which $\mathbf{X(t)}$  reaches a maximum, and, later, it decreases until it enters the point $\mathbf{P_{III}}$ . Conditions $\mathbf{bV < a f} $ and $\mathbf{h - \frac{1}{4e^2}<0}$    are satisfied.} 
	\label{phasevirus_fig3}
\end{figure}

\begin{align}
\label{equation13}
X_{crit} = \left( 1 + \frac{a}{\lambda_{2}^{(II)}}\right) X_2.
\end{align}
When $\bm{X_\infty}$  is small, the outcome can be very favorable.\\
For instance, when 
\begin{align}
\label{equation14}
X_\infty < \frac{11}{6e} + \frac{2f}{3d},
\end{align}
all the phase trajectories tend to the fixed point $\bm{P_{I}(X=0)}$.
\\
We can also apply the isocline method in order to further investigate the system. A careful analysis of the behavior of the phase trajectories allows us to conclude that the condition
\begin{align}
\label{equation15}
d > 4ef,
\end{align}
is favorable for the patient. This is a sufficient condition to avoid an uncontrollable rise of the virus population leading to the point $\bm{P_{IV}}$.
\\
In many cases, it is convenient to re-write the system \eqref{equation2} -\eqref{equation3} as one equation where the only unknown is $\bm{X(t)}$,
\begin{align}
\label{equation16}
\frac{d^2 X}{ d t^2} + \left[ f -d(X- eX^2) \right]\frac{d X}{ d t} - \frac{1}{X}\left( \frac{dX}{dt}\right)^2 = - \frac{d U(X)}{ dX}
\end{align}
In general, it is useful to discuss the dynamics of virus population as a general equation of the following type
\begin{align}
\label{equation17}
\frac{d^2 X}{d t^2} + F_{dis}\left(X, \frac{dX}{dt}\right)= - \frac{dU}{dX}
\end{align}
See Refs \cite{Ref16} for a simple explanation.
Equation \eqref{equation17} is equivalent to a Newton's equation for a “fictitious” particle moving in the potential $\bm{U(X)}$  under the action of nonlinear damping.

The potential  $\bm{U(X)}$  can have minima and maxima. So we can conceive the situation where the “fictitious” particle is trapped  inside a potential well. The particle needs to jump over a barrier for the virus population to continue increasing.
Studying the relative heights of the barriers, we get the condition
\begin{align}
\label{equation18}
9e(Vb -af) + 2 a d >0,
\end{align}
when this condition is satisfied, the “right” barrier of the potential well is higher than the “left” barrier. This case is more favorable for the patient.\\
A careful analysis shows that the condition

\begin{align}
\label{equation19}
2d > 9e f,
\end{align}
is very favorable for the patient.\\
The general meaning of conditions \eqref{equation15}, \eqref{equation18} and \eqref{equation19} is that the comparison between the values of $\bm{d}$ and the product $\bm{ef}$  can decide the outcome.\\
Let us analyze now the case $\bm{q\le 1}$. When
\begin{subequations}
	\begin{align}
	\label{equation20a}
	q\le 1,
	\end{align}                                                
the point $\bm{P_{I}}$  will be always unstable.
This means that it is almost impossible to reduce the virus population to zero.
This finding will play a very important part in the design of new therapies.
\begin{figure} 
	\centering
	\includegraphics[scale=0.8]{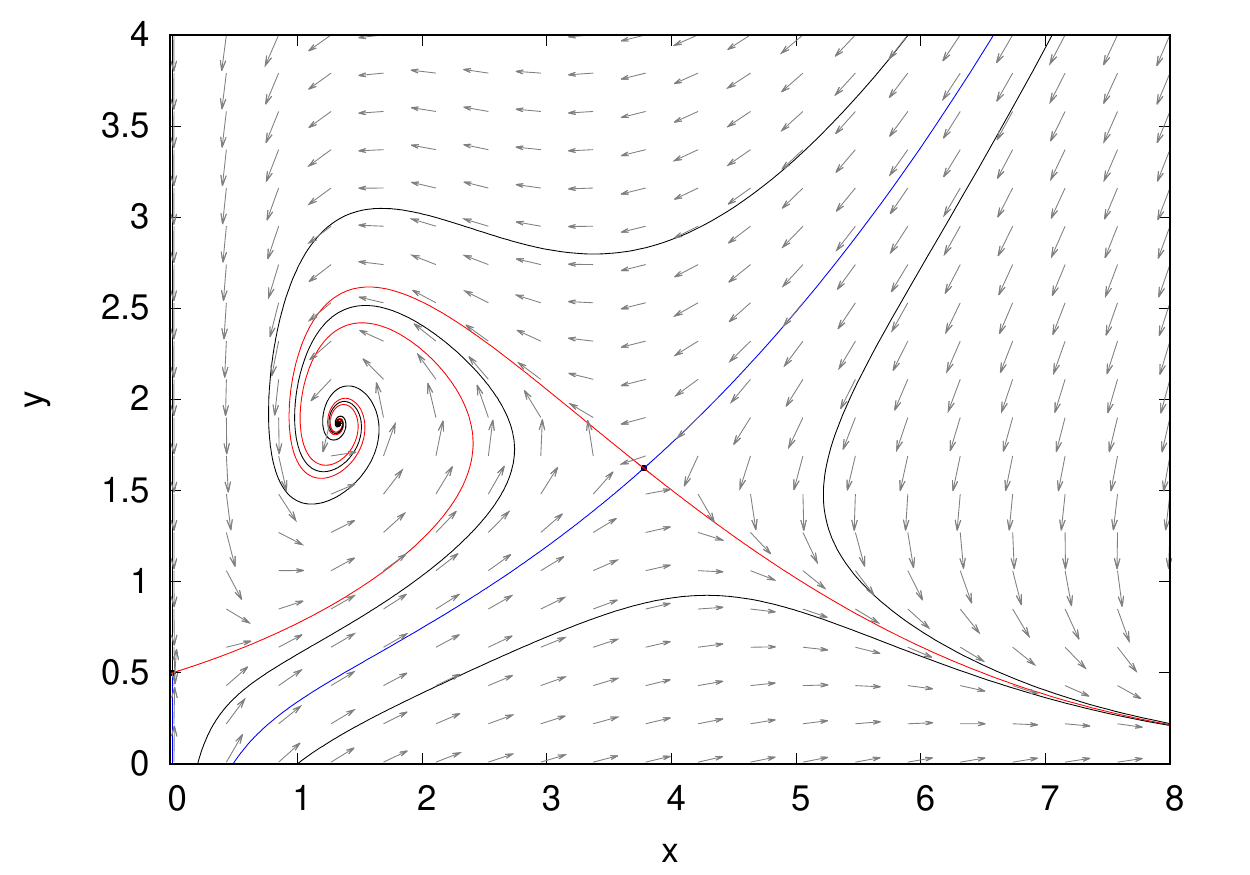}
	\caption{This dynamics is similar to that shown in  Fig. \eqref{phasevirus_fig3}. However, here the fixed point 
		$\mathbf{P_{III}} $ is a stable focus. } 
	\label{phasevirus_fig4}
\end{figure}
\\
Let us discuss time-dependent therapy against COVID-19.
\\
Let us consider the dynamical system \eqref{equation2}-\eqref{equation3} with time-dependent therapy \\
$\mathbf{c_1(t)}$  and $\mathbf{c_2(t)}=\epsilon \, \mathbf{ c_1(t)}$ where 
$\mathbf{\epsilon <1} $.
\\
Using ideas from \cite{Ref6},  we can obtain the following result.
If  $\mathbf{q<1}$, it is very difficult to cure the virus disease.
If we have a target decay for the virus population $\mathbf{X(t)}$ , then $\mathbf{ c_1(t)} $   must behave as 
\begin{align}
	c_1(t) =  \left[  \frac{ k X_\infty^{1-q}}{1-q} \right] \left( \frac{1}{X(t)^{1-q}}  \right) 
\end{align}
For instance, if we require the virus population to be reduced following a power law, say    
  $\mathbf{X(t)  \approx  \alpha/t^\gamma}$ , then the therapy must behave as $\mathbf{ c_1(t)} = t^{\gamma(1-q)}$. The exponent gamma represents the rate of decay of the virus population.
  \\
 We have designed therapies using the following late-intensification schedules:
\begin{align}
c_1(t)= c_0 \ln ( e + \delta t) \,\, \text{ and}  \\
 \,\,\, c_1(t)= c_0 \left[ 1 + \delta t \right]^{\gamma (1-q)}
\end{align} 
where $\mathbf{ c(t)=c_0} $  is a well-known constant-dose treatment taken by a patient for several days. (See Ref.   \cite{Ref4}).
Our logarithmic late-intensification schedule has been very successful  (See \cite{Ref5}
and references quoted there in). The traditional therapy is changed only slightly.
\\
However, the results are spectacular.

	\section{Monotherapies}
	\noindent
When $\bm{q>1}$, the parameters of the system and the initial conditions play an important role in the outcome. The virus-host interaction is decisive. There are situations where the immune system by itself can reduce the virus population to zero. Under other circumstances, the virus population will increase to numbers that can threaten the patients survival. If we apply conventional antiviral therapies with $\bm{c(t)=c_0}$ in the system \eqref{equation2} and \eqref{equation3} (where $\bm{c_{0}}$ is a constant), for $\bm{q>1}$, the cure can be accelerated \cite{Ref6}. 
	\\
If $\bm{q\le 1}$, then for any value of  $\bm{c_{0}}$, the virus population is never reduced to zero. The fixed point $\bm{P_{I}}$ is always unstable.
\\
The medical significance of this result can be expressed employing this statement: when $\mathbf{q\le 1}$ the disease is resistant to the immune response and the action of conventional therapy.


\begin{figure} 
	\centering
	\includegraphics[scale=0.8]{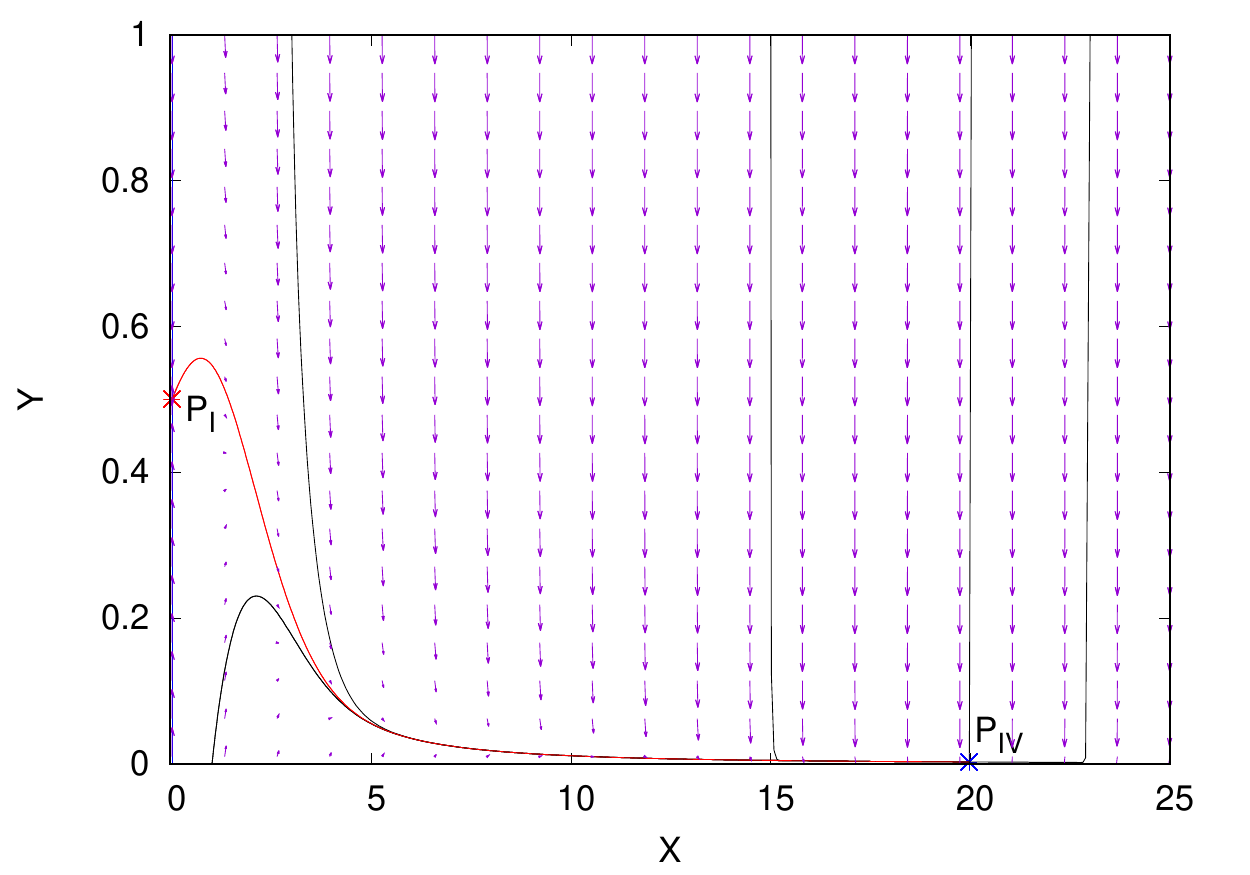}
	\caption{ In this case, the fixed points $\mathbf{P_{II}}$  and $\mathbf{P_{III}} $  disappear. Point $\mathbf{P_{I}}$  is unstable. All the phase trajectories tend to the point $\mathbf{P_{IV}}$ , where  $\mathbf{X \approx X_\infty} $. Conditions $\mathbf{af >Vb}$  and $\mathbf{h-\frac{1}{4e^2} >0}$  are satisfied. } 
	\label{phasevirus_fig5}
\end{figure}

	\section{Combination therapies }
	\noindent
	Our analysis shows that for $\bm{q\le 1}$, the virus can develop resistance both against the attack of the immune system and all conventional monotherapies with constant doses of the medication. All this investigation leads to combination therapies.
	\\
First, we have to use therapies that change the parameters in such a way that fixed point
$\bm{P_{I}}$  becomes asymptotically stable (a stable node). Then we need to apply therapies that will help the phase trajectory to go to the point $\bm{P_{I}}$.
The condition $\bm{q>1}$ should be completed with the stability of fixed point $\bm{P_{I}}$:

\begin{align}
	\label{equation20b}
	bV > \frac{ q k f}{q-1},
\end{align}

\end{subequations}
and complemented with condition \eqref{equation15}.
\\
This means that immuno-therapy is also very important for the development of antiviral therapies.
\\
This work can guide physicians to rationally design new drugs or a combination of already existing drugs for the development of antiviral therapies.
\\
\\
Condition \eqref{equation20b} shows that the killing ability of the immune system and the external flow of lymphocytes should be stronger than the virus replication and the natural death of immune system agents.
\\
\\
Additionally, condition \eqref{equation15} says that the reproduction of the lymphocytes should be stronger than the inhibition of the immune system due to the general health weaknesses created by the disease.
\\
\\
The perfect strategy is to use a therapy that can change $\bm{q}$  so that the fixed $\bm{P_{I}}$  can be, in principle, stable. Of course, this does not guarantee that the point $\bm{P_{I}}$ is stable.
\\
\\
The condition $\bm{q>1}$ is a necessary condition for the stability of point $\bm{P_{I}}$. 
However, it is not a sufficient condition.
\\
\\
Later we need another therapy that will change the other parameters (see section 3) so that the fixed point $\bm{P_{I}}$  is actually asymptotically stable.
This step is probably satisfied with an immuno-therapy.
\\
\\
Finally, we need a treatment $\bm{c(t)}$  that definitely kills the virus, leading the phase trajectory to the fixed point.
The ideal candidate for the first task could be a gene-targeted therapy.
On the other hand, we believe there are antivirals that can be utilized in order to accomplish this goal.
\section{Some biophysical analysis}
\noindent
Drug repurposing for SARS-CoV-2 is very important for our world. It can represent an effective drug discovery strategy from existing drugs. It could shorten the time and reduce the cost compared to de novo drug discovery \cite{Ref43}.
\\
\\
Phylogenetic analysis of 15 HCoV whole genomes reveal that SARS-CoV-2 shares the highest nucleotide sequence identity with SARS-CoV \cite{Ref43}. 
\\
\\
A molecular docking study has been published by Abdo Elfiky \cite{Ref44}.
\\
\\
The results show the effectiveness of Ribavirin, Remdesivir, Sofobuvir, Galidesivir, and Tenofovir as potent drugs against SARS-CoV-2 since they tightly bind to its RdRp. Additional findings suggest guanosine derivative (IDX-184), Sefosbuvir, and YAK as  top seeds for antiviral treatments with high potential to fight SARS-CoV-2 strain specifically.


\section{Real experiments and clinical studies}
We have reviewed the medical literature on COVID-19 treatments.
There is experimental evidence supporting combination therapies
\cite{Ref17,Ref18,Ref19,		Ref20,Ref21,Ref22,Ref23,Ref24,Ref25,Ref26,Ref27,Ref28,Ref29,
Ref30,Ref31,Ref32,Ref33,Ref34,Ref35,Ref36,Ref37,Ref38,Ref39,Ref40,
Ref41,Ref42,Ref43,Ref44,Ref44,Ref45,Ref46,Ref47,Ref48,Ref49,Ref50,
Ref51,Ref52,Ref53}
However, medical practice has been concentrated mostly on monotherapies. 
\cite{Ref17,Ref18,Ref19,		Ref20,Ref21,Ref22,Ref23,Ref24,Ref25,Ref26,Ref27,Ref28,Ref29,
Ref30,Ref31,Ref32,Ref33,Ref34,Ref35,Ref36,Ref37,Ref38,Ref39,Ref40,
Ref41,Ref42,Ref43,Ref44,Ref44,Ref45,Ref46,Ref47,Ref48,Ref49,Ref50,
Ref51,Ref52,Ref53}
\\
Even when combination therapies have been used, in many cases, the combinations have not been optimized. We believe we can improve the treatment outcomes using our results.
\\
Combination of antivirals is the most common therapeutic set. 
\cite{Ref17,Ref18,Ref19,		Ref20,Ref21,Ref22,Ref23,Ref24,Ref25,Ref26,Ref27,Ref28,Ref29,
	Ref30,Ref31,Ref32,Ref33,Ref34,Ref35,Ref36,Ref37,Ref38,Ref39,Ref40,
	Ref41,Ref42,Ref43,Ref44,Ref44,Ref45,Ref46,Ref47,Ref48,Ref49,Ref50,
	Ref51,Ref52,Ref53}
In many cases, the used antivirals were previously developed for other viruses (e.g. SARS, MERS, Ebola, Flu, and HIV).
\cite{Ref17,Ref18,Ref19,		Ref20,Ref21,Ref22,Ref23,Ref24,Ref25,Ref26,Ref27,Ref28,Ref29,
	Ref30,Ref31,Ref32,Ref33,Ref34,Ref35,Ref36,Ref37,Ref38,Ref39,Ref40,
	Ref41,Ref42,Ref43,Ref44,Ref44,Ref45,Ref46,Ref47,Ref48,Ref49,Ref50,
	Ref51,Ref52,Ref53}
\\
We present a summary of the studies about COVID-19 treatments.\\
Therapy 1: Lopinavir + Ritonavir (Antivirals against HIV) + Oseltamivir (Flu)
Lopinavir/Vitanovir are approved protease inhibitors for HIV.\\
Results: There is some scientific evidence that this combination can work
\cite{Ref17,Ref51,Ref52}
\\
Inconsistent results in some completed clinical trials.
\\
\\
Therapy 2: Antiflu Arbidol + Anti-HIV antiviral Darunavir.\\
Results: There is some scientific evidence that this combinations can help. \cite{Ref17,Ref51}
\\
\\
Therapy 3: Lopinavir/Ritonavir + Ribavirin.\\
This is an anti-HIV therapy used in SARS.\\
Results: There is some scientific evidence that this combination could work \cite{Ref18}.
\\
\\
Therapy 4: Remdesivir + Lopinavir/Ritonavir + Interferon beta.\\
Remdesivir interferes with virus RNA polymerases to inhibit virus replication, and was used for Ebola virus outbreak.\\
Results: The combination is being tested for MERS. There is some scientific evidence that this combination could work for other coronaviruses \cite{Ref18}
\\
\\
Therapy 5: Cloroquine, Hydroxycloroquine.\\
This is an antimalarial drug.\\
Results: Inconsistent results in completed clinical trials \cite{Ref2,Ref20,Ref21,Ref25,Ref26,Ref35}, and \cite{Ref38}.
\\
\\
Therapy 6: Hydrocloroquine + antibiotic azithromycin.\\
Results: Dedier Raoolt and coworkers published results of a completed clinical trial that “proved” efficacy \cite{Ref20}. However, nowadays, this therapy is considered controversial.
\\
\\
Therapy 7: Convalescent plasma.\\
Convalescent plasma from cured patients provides protective antibodies against \\SARS-CoV-2.\\
Results: Proven efficacy \cite{Ref23,Ref27,Ref28,Ref29,Ref30,Ref31,Ref32,Ref52}
\\
\\
Therapy 8: “Natural killer” cell therapy.\\
Natural killer cell therapy can elicit rapid and robust effects against viral infections through direct cytotoxicity and immunomodulatory capability.\cite{Ref52}\\
Results: Being tested in clinical trials \cite{Ref52,Ref30}, and \cite{Ref37}.
\\
\\
Therapy 9: EIDD-2801.\\
This is an antiviral.\\
EIDD-2801 is incorporated during RNA synthesis and then drives mutagenesis, thus inhibi-ting viral replication \cite{Ref52}.\\
Results: Being tested in clinical trials \cite{Ref36,Ref37,Ref52}.
\\
\\
Therapy 10: Remdesivir.\\
This is an antiviral that interferes with virus RNA polymerases to inhibit virus replication.\\
Results: Approved by FDA
Inconsistent and conflicting results in completed clinical trials. \cite{Ref39,Ref45,Ref46,Ref47,Ref52,Ref53}. This is a promising drug.
\\
\\
Therapy 11: Ivermectin.\\
Results: Leon Caly et al \cite{Ref48} have observed that the FDA-approved drug Ivermectin inhibits the replication of SARS-CoV-2 in vitro. The authors have shown that this dug actually “kills” the virus within $48$ hours.
This drug is being used extensively and massively in some countries (e.g. South America), in some cases, as a national policy.
However there are no completed clinical trials.
\\
\\
Therapy 12: Human monoclonal antibodies.\\
Results: Chuyan Wang et al have found a human monoclonal antibody $(47D11)$ that neutralizes SARS-CoV-2 \cite{Ref49} 
Being tested in clinical trials.
\\
\\ 
Therapy 13: Mesenchymal Stem Cells.
This is a cell therapy.
MSCs have regenerative and immunomodulatory properties and protect lungs against ARDS.\\
MSC therapy can inhibit the over activation of the immune system and promote repair improving the microenvironment. They regulate inflammatory response and promote tissue repair and regeneration \cite{Ref50}.
Results: Proven efficacy in completed clinical trials \cite{Ref52}.	
\\
\\
Therapy 14: Lopinavir/Ritonavir + Ribavirin + Interferon beta-1b1.\\
Results: Fan-Ngai Hung et al have published positive and promising results of a clinical trial\,\cite{Ref47}.


\section{New combination therapies}	
\begin{figure}
	\centering
	\includegraphics[scale=0.8]{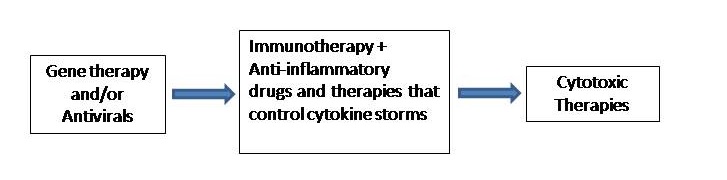}
	\caption{ General Therapeutic Plan } 
	\label{Figure1}
\end{figure}

We have estimated the parameters of the model (equations \eqref{equation2} and \eqref{equation3}) using published data from the dynamics of different kinds of biological populations
\cite{Ref17,Ref18,Ref19,Ref20,Ref21,Ref22,Ref23,Ref24,Ref25,Ref26,Ref27,Ref28,
	Ref29,Ref30,Ref31,Ref32,Ref33,Ref34,Ref35,Ref36,Ref37,Ref38,Ref39,Ref40,Ref41,Ref42,
	Ref43,Ref44,Ref45,Ref46,Ref47,Ref48,Ref49,Ref50,Ref51,Ref52,Ref53,Ref54,Wolfel,Pan,Munster,Bommer,Kim,To,Zheng,Lee,Zhou,Lescure,Zou,Corman,Vetter,Liting Chen}
\\
These data include sets virus population and cell populations. 
\\
In all the cases, the growth of the studied populations represents the most relevant behavior of the disease.
\\
Some examples are the virus population and the neoplastic cell population.
Normally, these are the only published data.
\\
The value of the virus population density is provided by the viral load = number-of-copies/mL. 
\\
Actually, what we usually know is $\bm{log_{10}}$ [number-of-copies/$\bm{mL}$, or $\bm{log_{10}}$ [number-of-copies/$\bm{1000}$ cells].
\\
Sometimes the parameters are estimated using best-fit solution functions. 
\\
In other cases, approximate values of the parameters are obtained from some characteristics of the dynamics like the slope of the tangent line to the graphed function, the fixed points, the stability conditions, and the eigenvalues of the Jacobian matrix.
\\
\\
Often, data about the immune behavior is not explicitly available. So, we use a version of the model that consists of one nonlinear differential equation only for $\bm{X(t)}$.
\\
However, that equation contains the parameters that characterize the immune system. Thus, these parameters can be estimated, too.
\\
We have observed several patterns in the virus dynamics\cite{Ref17,Ref18,Ref19,Ref20,Ref21,Ref22,Ref23,Ref24,Ref25,Ref26,Ref27,Ref28,
	Ref29,Ref30,Ref31,Ref32,Ref33,Ref34,Ref35,Ref36,Ref37,Ref38,Ref39,Ref40,Ref41,Ref42,
	Ref43,Ref44,Ref45,Ref46,Ref47,Ref48,Ref49,Ref50,Ref51,Ref52,Ref53,Ref54,Wolfel,Pan,Munster,Bommer,Kim,To,Zheng,Lee,Zhou,Lescure,Zou,Corman,Vetter,Liting Chen}
\\
\\
First pattern: the viral load increases rapidly and reaches a peak. Then the viral load declines due to the action of a strong immune system. The final viral load cannot be detected. We assume it is zero. 
(See Fig. \eqref{phasevirus_fig1}).
\\
\\
Second pattern: the viral load increases rapidly and reaches the peak, followed by a plateau. The plateau can be short or long. After the plateau, the viral load declines to zero. (In this case, the dynamics reaches a fixed point. Then, the parameters of the immune system change (e.g. $ \mathbf{b, V}$)).  Then the fixed point $\bm{P_I}$ is stable again.
\\
\\
Third pattern: the viral load increases rapidly and reaches a peak, followed by a plateau with a large value of the virus load. The plateau never ends. The patient dies. The viral load never declines. Fig. \eqref{phasevirus_fig5}.
\\
\\
Fourth pattern: the viral load increases rapidly and reaches a peak. Then the viral load declines. The decline is followed by a long plateau. The value of the virus load is much smaller than the peak. However it is far from zero.  (See Fig. \eqref{phasevirus_fig3})  
\\
These behaviors can also occur under the action of therapy \cite{Ref17,Ref18,Ref19,Ref20,Ref21,Ref22,Ref23,Ref24,Ref25,Ref26,Ref27,Ref28,
	Ref29,Ref30,Ref31,Ref32,Ref33,Ref34,Ref35,Ref36,Ref37,Ref38,Ref39,Ref40,Ref41,Ref42,
	Ref43,Ref44,Ref45,Ref46,Ref47,Ref48,Ref49,Ref50,Ref51,Ref52,Ref53,Ref54,Wolfel,Pan,Munster,Bommer,Kim,To,Zheng,Lee,Zhou,Lescure,Zou,Corman,Vetter,Liting Chen}
\\
\\
Let us introduce the units of the variables and parameters\\
Define the variable $\bm{X(t)}=\bm{log_{10}}\left[ \text{ number of copies / mL }\right] $.\\
$\left[\bm X(t)\right]= \bm {uv} $. Thus $\bm X=1\,  \bm {uv}$  if $(\text{the number of copies)}/mL =10$. Here $\bm{uv}$ stands for unit of viral load where $\bm{X}$ given in units of $\bm{log_{10}}$ [number-of-copies/mL].
\\
$\bm Y(t)$ is the number of lymphocytes/$mL$,
$\bm \left[\bm Y(t)\right]= 1 \, \bm{nc}$, where $1 \, \bm{nc}= 1 \,  \text{lymphocyte}/mL$,
$\left[\bm  a\right]= 1/day$, $\left[  \bm f \right]= 1/day$, $\left[\bm  b\right]= 1/(nc)day$,
$\left[\bm  V\right]= nc/day$, $\left[\bm  d\right]= 1/(uv)day$, $\left[\bm  a\right]= 1/uv$, $\bm q$ is dimensionless.
\\
Let us discuss some particular examples of real virus population growth.
\\
Example 1 (Patient 14 in Ref. \cite{Wolfel})
\\
$\bm{X_\infty} =8 $ uv, $\bm{k}=0.04$ (1/day), $\bm{q}=2$, $\bm{(a-Vb/f)}=-0.71$ (1/day),\\ $\bm{(d-4ef)}=0.6 $ (1/(uv)day).
\\
In this case, the immune system is so strong that it is able to eradicate the virus by itself. Virus load is approaching zero after $10$ days.
\\
\\
Example 2 (patient 7 in Ref. \cite{Wolfel})
\\
$\bm{X_\infty} =8.5 $ uv, $\bm{q}=2$, $\bm{k}=0.05$ (1/day), $\bm{e}= 0.1$ (1/uv), $\bm{b}=0.028$ (1/(nc) day),\\ $\bm{f}= 0.251 $ (1/day), $\bf{V}=0.002$ ((nc)/day), $\bm{a}=0.1$ (1/day), $\bm{d}=0.119$ (1/(uv) day).
\\
In this case, the viral load will reach the maximum. Then the virus load will decline. But it will not approach zero. The value of $\bm{X(t)}$ will be approximately constant for a long time.
\\
In the dynamical system this is a stable fixed point. The real data shows a long plateau where $\bm{2 uv < X < 3 uv}$. The known data does not show an end to this plateau.
\\
\\
Example 3 (patient 1 in Ref. \cite{Lescure})
\\
$\bm{X_\infty} =7.9 $ here $\bm{X}$ is given in units of $\bm{log_{10}}$ 
$[\text{number-of-copies}/1000$ cells]),
$\bm{q}=0.9$,\\ $\bm{k}=0.1$ (1/day), 
$\bm{f-Vb/a}= 0.25$ (1/day), $\bm{d-4ef}=0.2$ (1/(uv) day), \\
$(1/2e) - \sqrt{(1/4e^2)-h}=4.2$ uv$^{*}$. 
Initially, the viral load increases and reaches the maximum, followed by a plateau. Both the model and the real data agree with this.
\\
\\
Later this patient was treated with Remdesivir.
Now we re-estimate the parameters\\
$\bm{X_\infty} =7.3 $ uv$^{*}$, here $\bm{X}$ is given in units of 
$ \bm{log_{10}} [\text{number-of-copies}/1000$ cells]),  
$\bm{q}=2$, \\$\bm{k}=0.01$ (1/day), 
$\bm{a- Vb/f}= -4.2$ (1/day), $\bm{d-4ef}=0.3$ (1/(uv) day).
\\
Now both the model and the real data show that the viral load will approach zero!
\\
The antiviral changed parameters $q$ and $(a-Vb/f)$.
\\
The cases of patients $2, 4$ and $5$ from Ref. \cite{Lescure} are very similar to Example 1. 
\\
The immune system is able to eradicate the virus without external therapy.
\\
\\
Example 4 (patient 3 from Ref. \cite{Lescure})\\
$\bm{X_\infty} =7.8 $ uv$^{*}$, $\bm{q}=0.82$, $\bm{k}=0.04$ (1/day), 
$\bm{f- Vb/a}= 0.46$ (1/day),\\ $\bm{d-4ef}=-0.46$ (1/(uv) day).
\\
The viral load reaches a maximum, followed by a plateau.
This is an $80$ years old man with a very depressed immune system (he had had thyroid cancer).\\
This patient was sick with COVID-19 for $24$ days. He was medicated with Remdesivir starting on day $16$. 
\\
The viral load decreased slightly. However, the immune system was too weak.
\\
The viral load never reduced to zero. The patient died on day $24$.
This patient probably needed a combination therapy containing antivirals, immunotherapy and a virus-killing medication.
\\
\\
Example 5 (patient DF from Ref. \cite{Liting Chen})\\
$\bm{X_\infty} = 8.4 $ uv, $\bm{q}=0.6$, $\bm{k}=0.02$ (1/day), 
$\bm{b}=0.024$ (1/(nc)day), $\bm{d}=0.0001$ (1/(uv)day),
$\bm{f}=0.2$ (1/day),
$\bm{e}=0.02$ (1/(uv)),
$\bm{V}=0.0002$ ((nc)/day).
\\
The viral load increases until it reaches the maximum. The immune system is so weak that we can consider that it is not working at all.\\
The viral load will never decrease. The patient died on day $18$.
\\
\\
Example 6 (the case of the patient reported in Ref.\cite{Kim} )\\
$\bm{X_\infty} = 9.5 $ uv, $\bm{q}=1.8$, $\bm{k}=0.01$ (1/day), 
$\bm{a - Vb/f}= -0.35$ (1/day),\\ $\bm{d - 4ef}= 0.5$ (1/(uv)day).
\\
The viral load increases very fast and the peak is very high. Common sense would have led physicians to consider this case as critical. 
\\
\\
However, this patient was treated with a combination therapy.\\
According to the estimated parameters, we believe the treatment changed parameter $q$. 
\\
\\
The immune system was working well. The viral load is eradicated. This is seen in the dynamics of the model and in the real clinical data.\\
We have investigated all the data published in Refs
\cite{Ref17,Ref18,Ref19,Ref20,Ref21,Ref22,Ref23,Ref24,Ref25,Ref26,Ref27,Ref28,
	Ref29,Ref30,Ref31,Ref32,Ref33,Ref34,Ref35,Ref36,Ref37,Ref38,Ref39,Ref40,Ref41,Ref42,
	Ref43,Ref44,Ref45,Ref46,Ref47,Ref48,Ref49,Ref50,Ref51,Ref52,Ref53,Ref54,Wolfel,Pan,Munster,Bommer,Kim,To,Zheng,Lee,Zhou,Lescure,Zou,Corman,Vetter,Liting Chen}.\\
For instance, in Ref. \cite{Liting Chen}, the authors studied 52 patients. The cases are very similar to the examples and patterns that we have described here.
They found mild, severe, critical, and deadly cases.
\\
\\
In general, considering all the literature
\cite{Ref17,Ref18,Ref19,Ref20,Ref21,Ref22,Ref23,Ref24,Ref25,Ref26,Ref27,Ref28,
	Ref29,Ref30,Ref31,Ref32,Ref33,Ref34,Ref35,Ref36,Ref37,Ref38,Ref39,Ref40,Ref41,Ref42,
	Ref43,Ref44,Ref45,Ref46,Ref47,Ref48,Ref49,Ref50,Ref51,Ref52,Ref53,Ref54,Wolfel,Pan,Munster,Bommer,Kim,To,Zheng,Lee,Zhou,Lescure,Zou,Corman,Vetter,Liting Chen} here are interesting points that must be remarked. 
\\
\\
Some older patients with rapid evolution towards critical disease with multiple organ failure presented a long sustained persistence of SARS-CoV-2. \\
This persistent high viral load is explained by the ability of the SARS-CoV-2 to evade the immune response \cite{Lescure}.
\\
\\
 SARS-CoV-2 might be able to inhibit immune system signaling pathways, resulting in a malfunctioning of the immune system.\\
In most critical patients, the blood viral load was never eliminated \cite{Liting Chen}.
\\
\\
This can be explained with the stable fixed points of our model.\\
The results of our investigation of the model, the virus kinetics research, and the data from lab experiments and clinical studies
\cite{Ref17,Ref18,Ref19,Ref20,Ref21,Ref22,Ref23,Ref24,Ref25,Ref26,Ref27,Ref28,
	Ref29,Ref30,Ref31,Ref32,Ref33,Ref34,Ref35,Ref36,Ref37,Ref38,Ref39,Ref40,Ref41,Ref42,
	Ref43,Ref44,Ref45,Ref46,Ref47,Ref48,Ref49,Ref50,Ref51,Ref52,Ref53,Ref54,Wolfel,Pan,Munster,Bommer,Kim,To,Zheng,Lee,Zhou,Lescure,Zou,Corman,Vetter,Liting Chen}
 lead us to the following strategy to cure COVID-19:
\\
\\
A combination of antivirals can change the virus reproduction capabilities (parameter $\bm{k}$) and drug resistance (parameter $\bm{q}$).
This can make the fixed point $\bm{P}_{I}$ stable.
\\\\
A combination of immunotherapies can boost the immune system
 (parameters $(\bm{b, d, V})$.
\\
The agents of the immune system can reduce the virus load. 
(See  Figs. \eqref{phasevirus_fig1} -\eqref{phasevirus_fig4}).\\
A virus-killing therapy.\\
Even if the point $\bm{P}_{I}$ is stable and the immune system is working, it is possible that the virus dynamics is not riding a phase trajectory that is approaching the fixed point $\bm{P}_{I}$ , where $\bm{X} = 0$. For instance, if the initial condition is on the “right” of the separatrix of the saddle point $\bm{P}_{II}$, then $\bm{X(t)}$ is not approaching the point $\bm{X} = 0$.
\\
A virus-killing therapy can change the position of the initial point 
$\bm{(X_0, Y_0)}$, in such a way that this point will be on the “left” of the separatrix
 (See Fig. \eqref{phasevirus_fig1} ). 
\\
Now there is always a phase trajectory that will drive the viral load, $\bm{X(t)}$, to the point where $\bm{X(t)} = 0$.
The particular medications that will be used in every combination are selected from the set of drugs already tested in clinical trials. 
\\
\\
\noindent
The ideas discussed in the first $6$ sections of the paper lead to the conclusion that we need a combination therapy that contains at least some the following features:

\begin{enumerate}[label=(\Alph*)]
	\item A combination of drugs that impair somehow the biophysics of the virus replication, infection and/or treatment resistance.
	\item A combination of drugs that enhance the immune system ability to provide enough agents and their capability to fight the virus + anti-inflammatory drugs.
	\item A cell-killing therapy.
\end{enumerate}
Our paper is not only about mathematical  models. We have critically reviewed all the published data about possible medical treatments against COVID-19.
\\
\\
We have used a method that we have developed called Complex Systems Investigation to analyze the data.
\\
\\
Complex Systems Investigation contains ideas from Nonlinear Dynamical Systems, Inverse Problems, and Experimental Design Mathematics.
\noindent
Our results show that a successful treatment should be a combination of therapies as that shown in 
Fig. \eqref{Figure1}

\begin{figure}[hbt!]
	\centering
	\includegraphics[scale=0.8]{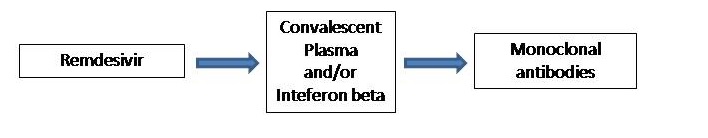}
\caption{One of the simplest therapies: Remdesivir plus Immunotherapy plus Monoclonal antibodies } 
	\label{Figure2}
\end{figure}

\begin{figure} 
	\centering
	\includegraphics[scale=0.8]{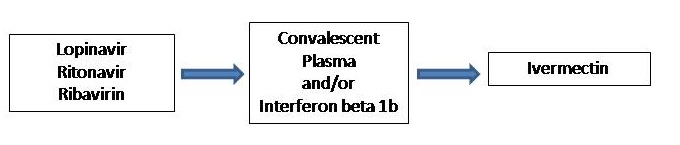}
	\caption{Anti-HIV cocktail plus Immunotherapy plus Ivermectin } 
	\label{Figure3}
\end{figure}

This is just a useful therapeutic plan.  We will see later that the role of a cytotoxic therapy sometimes can be played by an immunotherapy or an antiviral.  Fig. \eqref{Figure1} shows a very general plan.
\\
\\
Now we will present several concrete combination  therapies.  There are certain observations  \cite{Ref17} that support the existence of synergism between Remdesivir and monoclonal antibodies.
Considering the fact that our investigation leads to a combination of antivirals, immunotherapy, and virus – killing medications, the mentioned synergism help us build the treatment shown in  Fig. \eqref{Figure7}.
\\
\\
The simplest of our designed therapies is shown in Fig. \eqref{Figure2}   and  Figures
\eqref{Figure3}-\eqref{Figure10}
show different alternative treatments.

\begin{subfigures}
\begin{figure} [hbt!]
\centering
\includegraphics[scale=0.8]{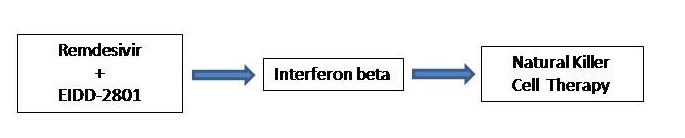}
\caption{Antiviral combination plus Interferon beta plus Natural Killer Cell Therapy } 
\label{Figure4}
\centering
\includegraphics[scale=0.8]{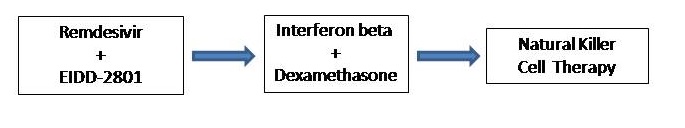}
\caption{Antiviral combination plus Interferon beta plus Natural Killer Cell Therapy plus Anti-inflammatory drug} 
\label{Figure4a}
\end{figure}
\end{subfigures}
\noindent
Baricitinib is an important anti-inflammatory drug. It has also anti-viral effects \cite{Ref54}.
\\
A commonly used steroid, Dexamethasone, can control the cytokine storms and can reduce the risk of death.
\\
Consider all the calculations presented together with the previous examples. We have done a similar research work with all the experimental data available in the references  
\cite{Ref17,Ref18,Ref19,Ref20,Ref21,Ref22,Ref23,Ref24,Ref25,Ref26,Ref27,Ref28,
	Ref29,Ref30,Ref31,Ref32,Ref33,Ref34,Ref35,Ref36,Ref37,Ref38,Ref39,Ref40,Ref41,Ref42,
	Ref43,Ref44,Ref45,Ref46,Ref47,Ref48,Ref49,Ref50,Ref51,Ref52,Ref53,Ref54,Wolfel,Pan,Munster,Bommer,Kim,To,Zheng,Lee,Zhou,Lescure,Zou,Corman,Vetter,Liting Chen}
\\
Sometimes, the data is very fragmented. In some cases, we only know the input, the medications, and the output.
\\
For instance, consider a patient with the following estimated parameters before therapy:
\\
$\mathbf{X_\infty}=9.3\, uv$, $\mathbf{q}=0.97 $, $\mathbf{k}=0.04 \, (1/day)$,
$\mathbf{b}=0.01\, (1/(nc)day)$, $\mathbf{d}=0.0005 \, (1/(uv)day)$, $\mathbf{e}= 0.03\, (1/(uv))$,
$\mathbf{V}=0.0001 ((nc)/day$, $\mathbf{X_0}= 4 uv$.\\
 Evidently, the patient has a bad prognosis. There is no way that this viral load will decrease under natural circumstances. We will apply the therapy shown in Fig. 
  \eqref{Figure5}.\\ 
Our result is the following:
\\
The first round (antiviral combination: Remdesivir + EIDD – 2801) will produce the parameters:
$\mathbf{q}=1.9$, $\mathbf{k}=0.01 \,(1/day)$. These are the only parameters that can be changed with the given antivirals. \\
After the immunotherapy (Convalescent plasma + Interferon beta), we get \\
$\mathbf{\left( a - Vb/f\right)}= - 0.82 \, (1/day)$ and $\mathbf{d- 4e f} = 0.71 \, (1/(uv)day)  $. \\
Additionally, the virus-killing medication (Natural Killer Cell Therapy) will reduce the "initial" viral load to the value 
 $\mathbf{ X_{02} < 1.4 uv <X_{crit}}$. Now there is a phase trajectory that can carry the viral kinetics to the stable fixed point $\mathbf{P_{I}}$  with the value $\mathbf{X}=0$. 
(See Fig.  \eqref{phasevirus_fig1}).\\
We can cure this patient with fulminant COVID – 19 infection!.
We believe these results can explain the clinical outcomes observed in references \cite{Ref26,Ref27,Ref28,Ref29}.

\begin{figure}
	\centering
	\includegraphics[scale=0.8]{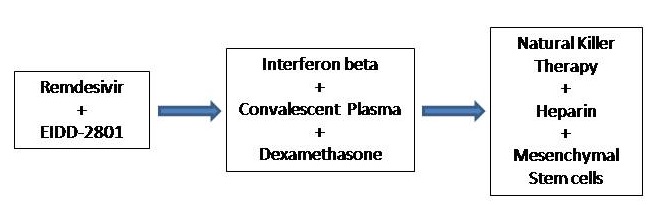}
	\caption{Powerful combination therapy that should kill the virus and save lives. The treatment includes a cocktail of antivirals: Remdesivir plus EIDD-2801, Immunotherapy, Natural Killer Cell therapy, Mesenchymal Stem Cells, a corticosteroid, and an anti-coagulant. } 
	\label{Figure5}
\end{figure}

\begin{figure}
	\centering
	\includegraphics[scale=0.8]{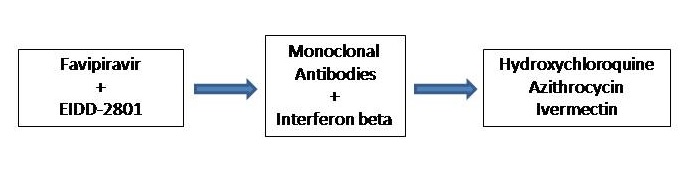}
	\caption{This is a combination of antivirals, an immune system booster, monoclonal antibodies, azithromycin, and the controversial hydroxychloroquine} 
	\label{Figure6}
\end{figure}

\begin{figure} 
	\centering
	\includegraphics[scale=0.8]{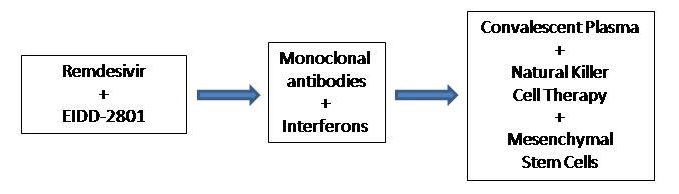}
	\caption{This therapy contains a combination of antivirals, monoclonal antibodies, an immune system booster, convalescent plasma, Natural Killer Cell Therapy, and an immune system modulator. This is a powerful combination.} 
	\label{Figure7}
\end{figure}
\newpage
\noindent
Regeneron pharmaceuticals has developed monoclonal antibodies to treat MERS. This company is already working on similar antibodies that might work against SARS-CoV-2.
\\
\\
Lopinavir/ritonavir + arbidol improved pulmonary computed tomography images \cite{Ref54}.
\\
\\
Interferons + Natural killer cells are promising. Interferons can enhance natural killer cells cytotoxicity.
Mesenchymal stem cells will act against inflammatory factors (cytokine storms).
\begin{figure} 
	\centering
	\includegraphics[scale=0.8]{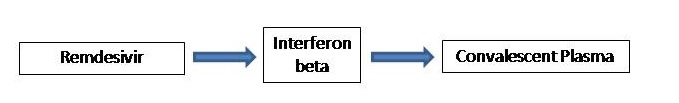}
	\caption{This is a next-door therapy. Any hospital should be able to provide this treatment, which could save patients' lives.} 
	\label{Figure8}
\end{figure}

\begin{figure} 
	\centering
	\includegraphics[scale=0.8]{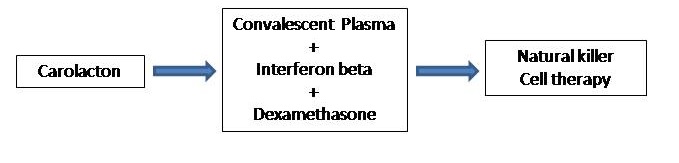}
	\caption{This could be a perfect realization of the General Therapeutic Plan (See Fig. \eqref{Figure1}): Gene therapy $\rightarrow$ Immunoterapy $+$ an  Anti-inflammatory drug $\rightarrow$ Cytotoxic therapy } 
	\label{Figure9}
\end{figure}

\begin{figure}[hbt!] 
	\centering
	\includegraphics[scale=0.8]{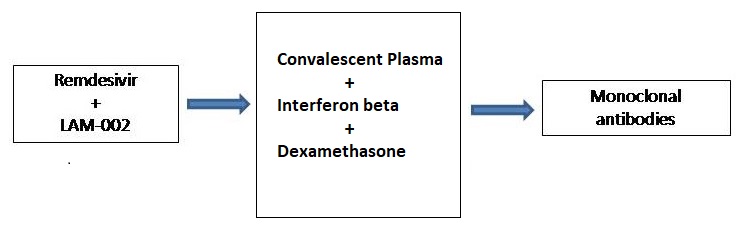}
	\caption{The combination Remdesivir plus LAM-002 should cripple the virus, the immunotherapies should kill the virus, and dexamethasone should relieve the inflammation and avoid the cytokine storms. The whole combination should control the immune system disorders, allergic reactions, and the breathing problems. This therapy should save lives and help patients recover faster.} 
	\label{Figure10}
\end{figure}
\noindent
MTHFV1 is a gene indispensable for viral replication in bat and human cells.
\\
\\
Carolacton is a MTHFV1 inhibitor. It is a natural bacteria-derived product \cite{Ref54}.
\\
\\
This is a good candidate for the first round in the combination therapy (see  Fig. \eqref{Figure9}).\\
A candidate for natural killer cell therapy is CYNK-001 \cite{Ref36}.
\\
\\
The most powerful therapy is shown in Fig.  \eqref{Figure5} . Probably this therapy should be used in the most severe critical fulminant cases.
\\
\\
On the other hand, Fig. \eqref{Figure8}  shows the next-door therapy. In principle, all elements should be available right now in every American city.

\section{Discussion}
\noindent
Remdesivir is considered the most promising drug for COVID-19 and MERS.
\\
\\
However, the clinical trials have produced conflicting results. Sometimes the results are encouraging, sometimes there are no significant benefits at all. Sometimes the people are still dying even taking remdesivir.
\\
\\
Our response to this paradox is that remdesivir will work as part of a combination therapy. Our result is that the idea of using remdesivir and some immunotherapies in combination would have profoundly excellent prospects. (See figures  \eqref{Figure1}-\eqref{Figure10}). 
\\
\\
We have tried to construct the combinations using drugs that have shown proven efficacy in completed clinical trials and/or laboratory experiments \cite{Ref54}.
\\
\\
Parameter $\bm{q}$ can be changed using drugs that change the nature of the virus.
\\
\\
Parameter $\bm{q}$ is related to the nature and structure of the virus.
\\
\\
For instance, the drug EIDD-2801 interferes with a key mechanism that allows the SARS-CoV-2 virus to reproduce in high numbers and cause infections.
\\
\\
EIDD-2801 is incorporated during RNA synthesis and then drives mutagenesis, thus inhibiting viral replication. So, this antiviral changes the nature of the virus.
\\
\\
Parameters $\bm{q}$ is related to the explosive reproduction of the virus and it is related to the difficulty to eradicate the virus.
\\
\\
The action of EIDD-2801 and Remdesivir is different. Remdesivir shuts down viral replication by inhibiting a key enzyme, the RNA polymerase.
\\
\\
Both Remdesivir and EIDD-2801 can change parameter $\bm{q}$.
Antivirals keep the virus from functioning and/or reproducing.
If we combine them, we can increase the probability that they will do the job of changing the biophysics of the virus. Then we can add immunotherapies to eradicate the virus.
\\
\\
These two antivirals are much more potent if given early.
In general, this is the case for most antivirals.
\\
\\
Some physicians can have concerns because, for them, it is not clear whether several combinations of medications and the high doses of the drugs in question could cause side effects.
\\
\\
Our research leads to the following solution to these problems: the addition of new drugs to the therapy and the total increase of doses can be administered using late-intensification schedules (e.g. logarithmic or power-law therapies \cite{Ref4,Ref5,Ref6}.
\\
\\
Our stable fixed point  with a small but finite virus population explains the following mystery: why a lot of patients who recovered from Coronavirus have retested positive \cite{Ref7}.
\\
\\
The existence of a finite minimum of the virus load in order to start an infection
(Eq. \eqref{equation13})  explains that there is a threshold value for a person exposure to sick people so that the person becomes infected. Our findings can also inform vaccine development. A vaccine works by training the immune system to recognize and combat viruses.
\\
\\ 
Some precedents.Therapy of HIV is complicated by the fact the HIV genome is incorporated into the host cell genome and can remain there in a dormant state for prolonged periods until it is reactivated. Some scientists believe that it is not possible to actually eradicate the virus completely.
\\
\\ 
Our research shows that this is a very striking example where $\bm{q \leq 1}$ . Following our ideas, it is possible that HIV can be completely eradicated. AZT was the first antiviral agent used for the treatment of HIV and was introduced in 1987. However, it became clear that mono therapy with AZT did not provide durable efficiency and hardly made any dent in the mortality rate.
\\
\\
Later, different studies showed that combination therapy with  two nucleotide analogues were better than monotherapy with only one.
\\
\\
After several experimental breakthoughs, a combination therapy known as HAART (highly active antiretroviral therapy) using two or three agents became available. By combining drugs that are synergistic, non-cross-resistant and no overlapping toxicity, it may be possible to reduce toxicity, improve efficacy and prevent resistance from arising.
\\
\\
All the antiviral drugs and therapeutic methods now known were discovered by random search in the laboratory.
\\
\\
We believe that using mathematical biophysics it is possible to create a rational approach for the discovery of new antiviral compounds and the design of the optimal combination therapy.

\section{Remarks}	
\begin{itemize}
\item We have developed a mathematical model to describe the SARS-CoV-2 viral dynamics. The model is a nonlinear dynamical system.
\item We have investigated the dynamical system theoretically and numerically.
\item 	We have found conditions for the stability of the fixed  point that corresponds to the complete eradication of the virus.
\item We identified the separatrix that separates the initial conditions that lead to the maximum value of the viral load from the initial conditions that lead to a limited growth of the virus population.
\item 	We have studied the global dynamics of the dynamical system. We can predict the evolution of any initial condition.
\item The fixed point $\mathbf{X =0}$ is stable when 
\begin{align}
q>&1,\\
Vb >& a f
\end{align}
\item If the following conditions are satisfied
\begin{align}
	af > Vb,&\\
	h - \frac{1}{4e^2}& >0,
\end{align}
then the separatrix does not exist and there are no restrictions to the growth of the viral load. This is a terrible situation.
\item 	Furthermore, condition $\mathbf{q\le1}$  means that the virus cannot be eradicated by the immune response or using any conventional monotherapy.
\item 	Let us discuss the biological meaning of the following conditions
\begin{align}
\label{eq28}	Vb > af ,\\
\label{eq29}	d > 4 ef ,\\
\label{eq30}	q>1.
\end{align}
In the real-life scenario, conditions \eqref{eq28}-\eqref{eq30}  mean that the immune system is working well and the virus infection is not drug resistant.
The combination therapy must be able to generate conditions \eqref{eq28}-\eqref{eq30}.
\item Our study provides explanations to several phenomena that have been observed during the experimental studies of SARS-CoV-2 virus.
\item 	We have critically reviewed the experimental and clinical literature about COVID-19.
\item 	Using the results from the investigation of the model and experimental data from laboratory and clinical studies, we have designed new combination therapies against COVID-19.
\end{itemize}

\section{Conclusions}
\noindent
J. H. Bergel et al \cite{Ref54} have published a paper in The New England Journal of Medicine with the information about the NIAID-supported study titled “Remdesivir for the treatment of COVID-19”. 
\\
\\
NIAID director had said that remdesivir will become the standard care of COVID-19.
\\
\\
The drug shortened the course of illness from an average of 15 days to about 11 days.
\\
\\
However, it is clear that the drug is not enough to help patients.
\\
\\
The medication is not a cure and it does not act quickly.
There is high mortality despite the use of remdesivir. So, remdesivir is not sufficient to cure patients. 
\\
\\ 
It seems that remdesivir does not cause an excess of side-effects.
\\
\\
Our take is that remdesivir alone is not enough.
Many other treatments, given as monotherapies, have failed to provide the promised results.
\\
\\
Our conclusion is that we need new scientifically designed combination therapies.
\\
\\
Using mathematical models and experimental data from laboratory  and clinical studies, we have been able to design new therapies, which, we expect, will cure the patients. (See figures \eqref{Figure1}-\eqref{Figure10}).
\\
\\
The new therapies also should be validated in double-blind, placebo-controlled trials with a large number of patients. 

\section{DATA AVAILABILITY}
The data that supports the findings of this study are available within the article.


\end{document}